\documentclass[10pt,letterpaper]{article}
\usepackage[top=0.85in,left=2.75in,footskip=0.75in]{geometry}

\usepackage{amsmath,amssymb}

\usepackage{changepage}

\usepackage[utf8x]{inputenc}

\usepackage{textcomp,marvosym}

\usepackage{cite}

\usepackage{nameref,hyperref}

\usepackage[right]{lineno}

\usepackage{microtype}
\DisableLigatures[f]{encoding = *, family = * }

\usepackage[table]{xcolor}

\usepackage{array}

\newcolumntype{+}{!{\vrule width 2pt}}

\newlength\savedwidth

\newcommand\thickhline{\noalign{\global\savedwidth\arrayrulewidth\global\arrayrulewidth 2pt}%
\hline
\noalign{\global\arrayrulewidth\savedwidth}}

\raggedright
\usepackage[aboveskip=1pt,labelfont=bf,labelsep=period,justification=raggedright,singlelinecheck=off]{caption}

\makeatletter
\renewcommand{\@biblabel}[1]{\quad#1.}
\makeatother

\usepackage{lastpage,fancyhdr,graphicx}
\usepackage{epstopdf}
\fancyheadoffset[L]{2.25in}
\fancyfootoffset[L]{2.25in}
\mathchardef\mhyphen="2D

\usepackage{rotating}
\usepackage{xcolor}
\newcommand\hl[1]{%
  \bgroup
  \hskip0pt\color{black}
  #1%
  \egroup
}

\begin{document}
\vspace*{0.2in}

% Title must be 250 characters or less.
\begin{flushleft}
{\Large
\textbf\newline{Ready Student One: Exploring the predictors of student learning in virtual reality} % Please use "sentence case" for title and headings (capitalize only the first word in a title (or heading), the first word in a subtitle (or subheading), and any proper nouns).
}
\newline
% Insert author names, affiliations and corresponding author email (do not include titles, positions, or degrees).
\\
J. Madden\textsuperscript{1*},
S. Pandita\textsuperscript{2},
J. P. Schuldt\textsuperscript{2},
B. Kim\textsuperscript{2},
A. S. Won\textsuperscript{2},
N. G. Holmes\textsuperscript{3},
\\
\bigskip
\textbf{1} Astronomy and Space Sciences, Cornell University, Ithaca NY, USA
\\
\textbf{2} Communication, Cornell University, Ithaca NY, USA
\\
\textbf{3} Laboratory of Atomic and Solid State Physics, Cornell University, Ithaca NY, USA
\\
\bigskip

% Use the asterisk to denote corresponding authorship and provide email address in note below.
* jmadden@astro.cornell.edu

\end{flushleft}

\nolinenumbers

% Please keep the abstract below 300 words
\section*{Abstract}
Immersive virtual reality (VR) has enormous potential for education, but classroom resources are limited. Thus, it is important to identify whether and when VR provides sufficient advantages over other modes of learning to justify its deployment. In a between-subjects experiment, we compared three methods of teaching Moon phases (a hands-on activity, VR, and a desktop simulation) and measured student improvement on existing learning and attitudinal measures.  While a substantial majority of students preferred the VR experience, we found no significant differences in learning between conditions. However, we found differences between conditions based on gender, which was highly correlated with experience with video games. These differences may indicate certain groups have an advantage in the VR setting. 

\nolinenumbers
\section*{Introduction}
In twenty-first century education, technology is pervasive in our classrooms~\cite{Maddux2003}. Research has found many ways in which technology benefits student learning and attitudes towards science~\cite{DBER2012}. As new instructional technologies are developed, it is necessary that researchers conduct critical evaluations of their effectiveness. There are many open research questions related to identifying how to most effectively use different kinds of technology for different learning goals or for different points in the learning process. One technology that has received particular attention for its potential in science learning is virtual reality and other immersive media \cite{dede2009immersive, pan2006virtual}.

In a typical college lecture for a science course, an instructor can choose to engage students with a concept using technologies such as specialized equipment for an interactive lecture demonstration~\cite{Sokoloff1997}, a dynamic computer simulation~\cite{Wieman2008}, or classroom polling with personal response systems~\cite{Stains2018}. Science courses also employ technology extensively in instructional labs, where students can use technology to obtain first-hand experiences with the phenomenon~\cite{Hofstein2004}. Instructional labs, however, have failed to provide measurable gains in student learning of those phenomena~\cite{Holmes2017, Etkina2010}, reminding us that what is important for learning is not the technology itself, but how and why it is used~\cite{DBER2012}. For example, using relatively similar equipment, interactive lecture demonstrations have consistently found measurable learning gains when students are actively engaged in making predictions and explaining the observations from the demonstrations~\cite{Sokoloff1997, Crouch2004}. Interactive simulations have also been shown to improve or replicate learning compared to hands-on manipulatives, while reducing the associated resources~\cite{Finkelstein2005, Chini2012}. While there are many potential explanations for these differences~\cite{SmithDemos}, two relate to how learners and technology interact through issues of embodiment and real-world complexities. This study aimed to test the impact of these variables by directly comparing student learning and attitudes from three different instructional technologies (a hands-on activity, desktop simulation, and virtual reality simulation), while taking advantage of their respective affordances along the dimensions of embodiment and real-world complexity.

\subsection*{Embodiment}
Theories of learning argue that cognition is inherently embodied: ``the mind must be understood in the context of its relationship to a physical body that interacts with the world.''~\cite[p.625]{Wilson2002}. Research has found that learning benefits from activities that explicitly attend to embodied cognition~\cite{ANDERSON200391, Roth2013,Carbonneau2013}. There are several ways in which embodiment is argued to support learning, generally tied to a hypothesis whereby activities help move cognition from abstract to concrete representations of a phenomenon~\cite{ANDERSON200391}.

Two such notions of embodied cognition focus on how learners off-load cognitive work on to the environment and how off-line cognition is body-based~\cite{Wilson2002, Martin2005}. These notions suggest physical aspects of cognition, whereby learning is supported through engaging perceptuo-motor systems~\cite{Tsang2015}. Indeed, nearly all science education advocates for the use of interactive hands-on activities~\cite{Ruby2001}.  Through hands-on activities and demonstrations, learners connect abstract concepts to their physical environment. For example, researchers in physics education have developed embodied activities for teaching concepts of energy conservation~\cite{Scherr2013}. In these activities, learners assign units of energy to physical objects (either cubes or people)~\cite{Scherr2012}, and then manipulate those objects to represent processes of energy transfer and dynamics. Through these concrete representations of an otherwise abstract phenomenon, learners develop their conceptual and mechanistic understandings of the phenomenon~\cite{Scherr2013}. By manipulating physical objects, students can see deep features of the phenomena, allowing them to effectively integrate the features into their mental models of the phenomena~\cite{Tsang2015, Piaget2013}.

Two other notions of embodied cognition focus on how cognition is situated within the real-world and that the learning environment is part of the cognitive system~\cite{Wilson2002, Martin2005}. Because the knowledge learned is tied to the environment in which it was learned, it is argued that there are limitations to applying the knowledge in new environments~\cite{Brown1989}. In science education, the initial learning environment can often seem too far removed or distinct from the intended environment for applying the knowledge. Context-rich activities, can provide concrete, real-world scenarios for otherwise abstract learning activities. Alternatively, activities may be modified to better represent the real world, such as by providing realistic representations of the phenomena.

These activities also demonstrate that there are degrees of embodiment. These may range from metaphorical groundings of cognition (for example, being able to imagine oneself walking from point A to point B) to physically experiencing the phenomenon (for example, actually walking from A to point B)~\cite{ANDERSON200391}. 
While much research has demonstrated the benefits of physically experiencing the phenomenon through hands-on activities, the results are not universal~\cite{Cunningham1946, Holmes2017, Chini2012, Finkelstein2005,Klahr2007}. One explanation for the lack of clear benefit for hands-on activities is that when the activities and materials are complicated or difficult to manipulate, the learner may experience extraneous cognitive load~\cite{Paas2003} or may be likely to make mistakes~\cite{Sokoloff1997}.
That is, real-world complexity gets in the way of student learning in hands-on activities. 

\subsection*{Real-world complexity}
In a real-world experiment, students’ measurements and observations are prone to variability, systematic effects, and measurement mistakes that are not relevant to the theoretical concepts being taught. In general, people struggle to evaluate uncertain events~\cite{Kahneman1972} and use error-prone shortcuts and heuristics in making judgments under uncertainty~\cite{Tversky1974}. Learning an underlying concept, then, becomes difficult if the information being used to develop that understanding is uncertain or probabilistic.

For hands-on activities, the uncertain and probabilistic elements contribute to extraneous cognitive load during the activity. Cognitive load refers to a learner's capacity for processing information in their short-term memory~\cite{Paas2003}. Information can either contribute to extraneous (ineffective) cognitive load or germane (effective) cognitive load~\cite{Paas2003}. Germane cognitive load refers to information that is relevant for learning, such that large amounts of germane cognitive load can improve learning~\cite{Kapur2016}. Extraneous cognitive load refers to information that is irrelevant for learning, such that large amounts of extraneous cognitive load impede learning~\cite{Sweller1991,Paas2003}. In hands-on activities, issues of uncertainty, complicated equipment, and user mistakes contribute to extraneous cognitive load, which may hamper learning. 

To reduce that cognitive load, hands-on activities often use heavily guided instructions that attempt to limit mistakes, tell students how to manipulate the equipment, and reduce uncertainties~\cite{Hofstein2004}. However, constructivist theories of learning argue that students must have the ability to explore and generate their own knowledge~\cite{Bransford1999}. The key is to develop activities with high germane cognitive load, but low extraneous cognitive load~\cite{Kapur2016, Schwartz2011}.

Simulations are one way to remove the extraneous cognitive load of real-world hands-on activities, allowing phenomena to be demonstrated in a consistent and controlled way. With that control, they can still be relatively unstructured, maintaining high germane cognitive load. Students can autonomously and easily change variables, allowing them to learn at their own pace~\cite{Price2018, podolefsky2010}. Several studies have found that computer simulations produce equal~\cite{Darrah2014, Chini2012, Evangelou2018} or better~\cite{Finkelstein2005, Chini2012} learning than hands-on activities. Simulations also provide opportunities for students to see features of a phenomenon that they would be unable to see otherwise, for example abstract concepts such as heat~\cite{Strzys2018} or microscopic objects such as cells~\cite{Chang2016} or electrons~\cite{Finkelstein2005, Kapp2019}.

Simulations, however, provide a more limited embodied cognition experience than hands-on activities, where the learner interacts directly with the phenomenon.
Virtual reality is a potential technology that can employ high-levels of embodiment, while maintaining controlled and simplified representations of the phenomena to be learned.

\subsection*{Why virtual reality?}
Immersive virtual reality (VR) may provide the best of both worlds. VR allows embodied simulations and offers a number of other affordances~\cite{Bricken1991, Perone2016} that make it uniquely suited as a teaching tool for basic science. First, students can physically interact with content, providing the engagement of a hands-on activity but with the control and replicability of a simulation.  Second, the simulations provide multiple forms of embodiment, such as changing perspectives to experience phenomena as they would in different circumstances in the real world~\cite{ANDERSON200391}. Third, students can experience these perspectives in ways unavailable in the real world~\cite{Price2018, Strzys2018, Kapp2019}. From a research perspective, the ability to track student movement allows for assessing engagement and learning~\cite{Won2014} to better test the embodiment hypothesis. It also facilitates implementing interventions to increase learning in real time. 

There is thus a need for experimental research that directly tests the effectiveness of VR on science learning, over-and-above that offered by existing hands-on and simulation approaches. Several studies have compared learning between these three modalities and found that, in terms of student attitudes, participants generally prefer learning in VR over other modalities~\cite{Smith2017, Chang2016}. One study even found that students' attitudes towards socio-scientific issues improved more in an augmented reality (AR) simulation over a desktop simulation, even though these aspects were secondary to the activity's primary cognitive goal~\cite{Chang2016}. Related to embodiment, several studies have found that participants became more immersed in VR than in other environments~\cite{Lier2018, Makransky2017, Winn2002}. Generally, this immersion seems to be independent of personal characteristics, such as gender, VR experience, and time spent gaming~\cite{Lier2018}. 

Measures of learning from VR are generally conflicting. Studies have found that participants in VR learn more than~\cite{Winn2002, Strzys2018}, as much as~\cite{Smith2017, Chang2016}, or less than~\cite{Makransky2017} participants in hands-on or desktop conditions. When learning was improved in VR over a hands-on activity, the gains were attributed to 
immediate feedback available through the simulation and visualization of the abstract phenomenon that was otherwise imperceptible in the hands-on activity~\cite{Strzys2018}. 
When learning was hindered in VR over a desktop simulation, the differences were attributed to 
higher cognitive load~\cite{Makransky2017}. In this study, where participants answered multiple-choice questions and performed technical lab procedures in the two types of simulation, researchers found that students in the desktop simulation condition learned more than students in the VR condition on a conceptual test. However, learning was the same between conditions on a transfer test and participants overwhelmingly preferred the VR condition. The researchers also found that participants in the VR condition had significantly higher cognitive load as measured through an electroencephalogram (EEG). They suggested that the physical manipulation of the equipment in VR was more complicated than the desktop condition, which may have increased students' extraneous cognitive load, impacting learning. They recommended experiments that used more natural control systems to manipulate the environment. 
Furthermore, the researchers also argued that the enjoyment associated with VR actually distracted the learners from learning. 

In addition to cognitive load, there are also questions about gender differences and experience with 3D rotations through experience with video games. In one study, men outperformed women during the task itself, but there was no difference on a post-condition recall test~\cite{Leon2018}. In a study that found no overall differences in learning from VR compared with video and static images, the researchers found that men and participants with experience with video games outperformed in the VR condition over other conditions and participants~\cite{Smith2017}. 
Because gender was also correlated with video game experience, they hypothesized that video game experience was a proxy for the gender differences emerging in their study.
This result is somewhat surprising given that neither condition involved much interaction with the simulations (even the VR participants could only look around the simulation). 

Research has found, however, that performance is improved when participants can more fully interact with the simulation: for example, walking around the simulation compared with remaining stationary while looking around the simulation~\cite{Leon2018}. Furthermore, one study found that students' reported sense of presence in a simulation was correlated with their learning from the simulation~\cite{Winn2002}. This study also found that preferential learning from VR was confined to sub-topics involving ``dynamic three-dimensional processes, but not processes that can be represented statically in two dimensions"~\cite[p.1]{Winn2002}. This suggests that VR simulations are more effective when they take advantage of their specific affordances.

The existing research exemplifies the idea that it is not the technology, but how it is used, that promotes learning. In this study, we aimed to develop and test a VR simulation that took advantage of its various affordances, particularly related to embodiment and real-world complexity. We probed these ideas by comparing an interactive VR condition with analogous hands-on and desktop simulation activities for learning about Moon Phases. As per the previous research, we evaluated students' conceptual knowledge, long-term retention, attitudes towards the activity, and socio-scientific beliefs. We also compared differential effects on sub-populations of our participants, including evaluating effects for gender, video game experience, and experience in VR. We found that there were no overall differences in learning on short-term or long-term assessments between conditions and that the immersive VR did not impact students' socio-scientific beliefs. We also replicated previous work in that participants overwhelmingly preferred the VR condition, and that men outperform in the VR condition, which may be attributed to video game experience.

\section*{Methods and Materials}
This study had two hypotheses. First, we proposed that there would be a main effect of virtual reality on learning. Second, we proposed that there would be corresponding effects on environmental attitude. We also collected data on other measures in order to explore interactions that might be predicted by the literature. All measures are reported below.

\begin{figure}[!ht]
\includegraphics[width=\columnwidth]{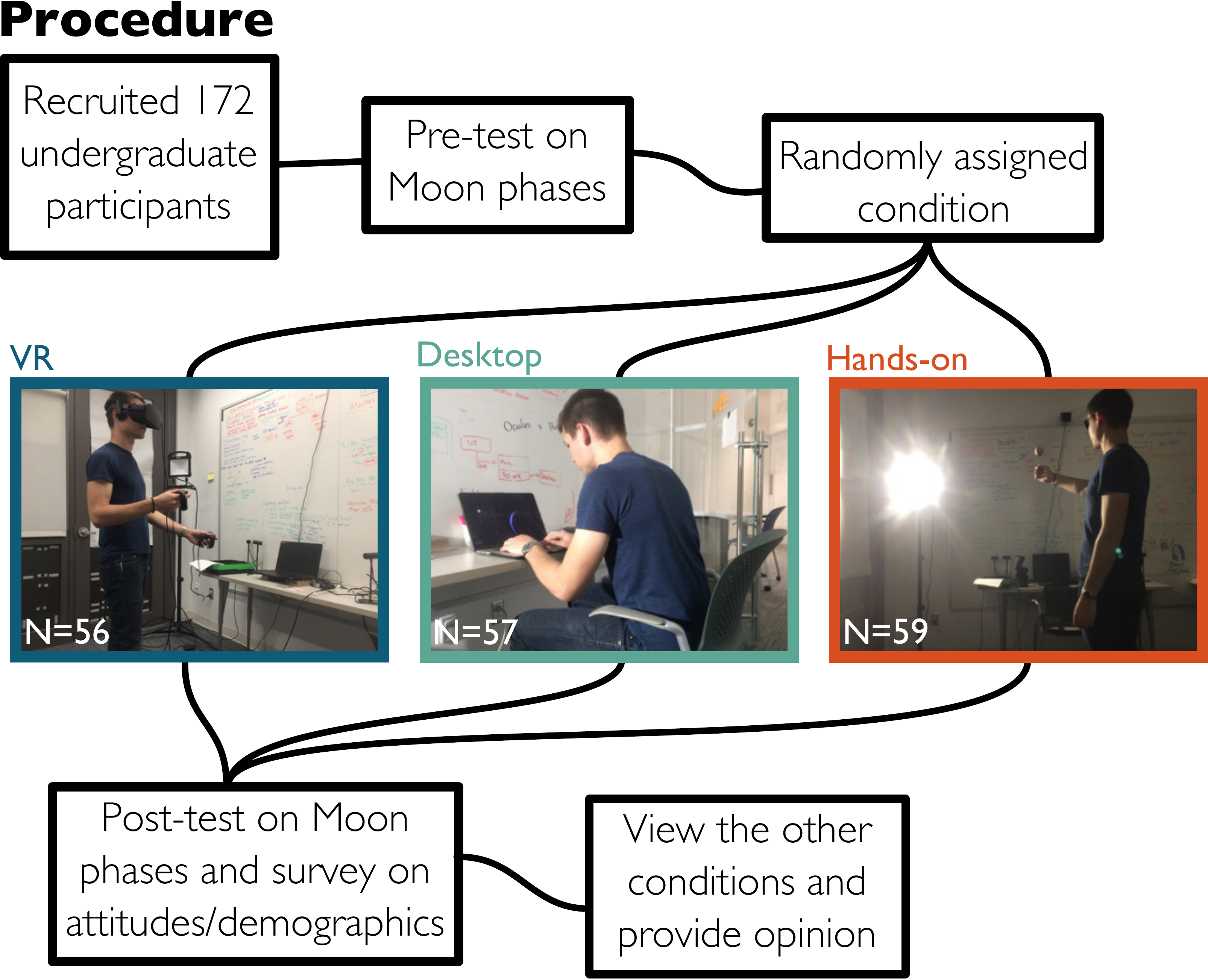}
\caption{{\bf Experiment procedure}
Map of the process for our experiment showing the three conditions.\emph{The individual shown has given written consent to have their likeness presented here.}}
\label{fig:flowchart}
\end{figure}

In our study, we used a between-subjects pre-post design, with three conditions. The three conditions were designed to express the same content using different educational tools; an immersive simulation using a VR headset, a computer-based interactive desktop simulation, and an analog hands-on activity (Fig. \ref{fig:flowchart}). We chose the concept of Moon phases, as it was expected to benefit from embodiment and reduction of real-world complexities. A review of over 35 years of astronomy education literature found that phases of the Moon was one of the most challenging topics in astronomy education~\cite{Lelliott2010}. A lesson on Moon phases requires the student to place themselves in spatial and temporal perspectives of the Sun-Earth-Moon system that are generally inaccessible, which can be challenging through static images or text~\cite{Galano2018, Turk2015}. Previous work has found that students' spatial reasoning correlates with their understanding of lunar concepts~\cite{Wilhelm2013, Cole2018}. Understanding of Moon phases also requires understanding the dynamic evolution of the phases over time and space~\cite{Cole2018}, likely facilitated through interacting with and moving around a simulation.  Furthermore, the traditional hands-on activity for teaching Moon phases (described below) is susceptible to various real-world inaccuracies, such as creating eclipses every month or rotating or orbiting the wrong way. Research has also found that explanatory features of traditional descriptions and images can interfere with student understanding~\cite{Galano2018, Turk2015}, motivating the need for authentic real-world visualizations over abstracted ones.

    \subsection*{Participants}
Participants were recruited from the undergraduate student population of a medium-sized private university. There were 172 participants, including 138 women, 31 men, and 3 other, all between the ages of 18 and 24. For three conditions to obtain a Cohen's d effect size of 0.25 with 80\% statistical power at 95\% significance level, our power test determined that a sample size of 53 participants per condition (159 total) would be sufficient. We ran a total of 172 participants to buffer against potential exclusions. The study primarily drew from students currently enrolled in an introductory astronomy class (varying in major), and students majoring in Communication. Participants' self-reported race and ethnicity were as follows: 62 Asian/Pacific Islander, 59 Caucasian, 21 African American, 18 bi/multiracial (including Asian/Pacific Islander, Caucasian, African American, Hispanic/Latinx, and Native American), 6 Hispanic/Latinx, 4 preferred not to answer, and 2 specified `other' without elaboration. Participants could select as many categories that applied. All participants signed an informed-consent form before beginning the experiment. All aspects of the experiment were approved by the Cornell Institutional Review Board Protocol \#1708007381. Participants were compensated with course credit or 15 dollars in cash for their participation. The individual pictured in Fig. \ref{fig:flowchart} has provided written informed consent (as outlined in the PLOS consent form) to publish their image alongside the manuscript.

Participants first took a pre-test and then they were randomly assigned to one of three conditions. After the activity (which included consistent self-guided lessons on Moon phases, described below), participants took a post-test that included a demographic and attitudinal survey. Finally, the participants were shown the other two conditions and asked to comment on which they would prefer as their favorite learning method and why. These activities are summarized in Fig. \ref{fig:flowchart}. A large preference for one condition over another may lead to greater retention of knowledge learned in that condition. To explore this potential effect we contacted all participants four months after participating and asked them to complete another learning test (delayed post-test). We received 56 responses to the delayed post-test making a breakdown by demographics difficult.

    \subsection*{Conditions}
Each condition was designed to give participants a similar learning experience using the three technologies we employed. The overall design was to recreate a Sun-Earth-Moon system that the participant could control in time in order to observe the changes in the Moon's phase and the positions of the Sun, Earth, and Moon during each phase. In each condition the participant could move forward and backward in time and had control over the perspective from which the system was being viewed. Each condition also contained guiding questions for the participant to assist with navigating and learning from the simulation. 

        \subsubsection*{VR Simulation}
\begin{figure}[!ht]
\includegraphics[width=0.8\columnwidth]{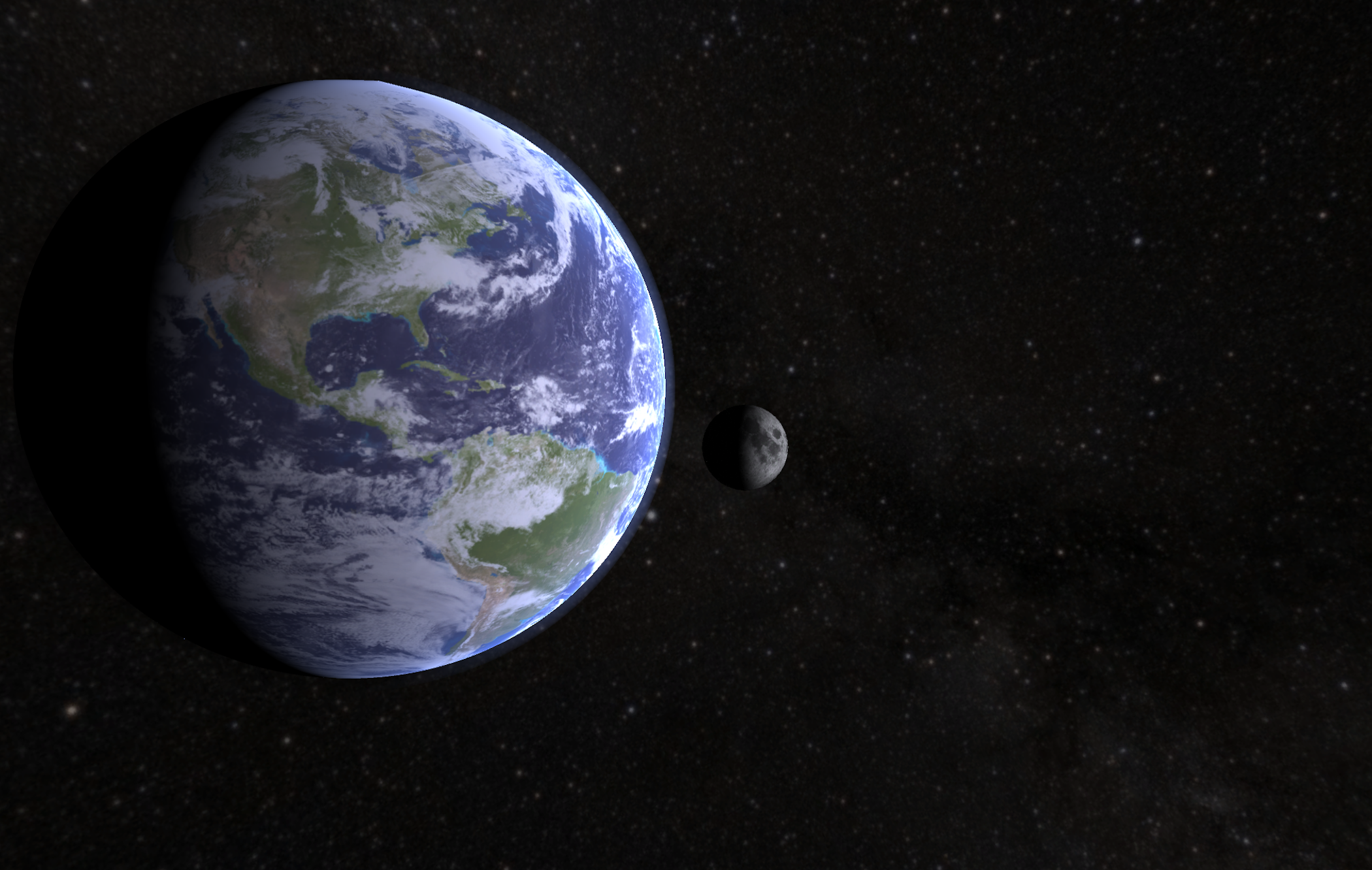}
\caption{{\bf Simulated activity}
A screenshot taken from inside the simulated Moon phases activity showing what the VR and Desktop conditions looked like. A video of the VR experience can be found online: \url{https://vimeo.com/310212130} }
\label{pic:earthmoonpic}
\end{figure}

The VR simulation was designed to mimic the hands-on activity as closely as possible, while still taking advantages of the technology's unique affordances. In the VR simulation, participants used a headset and controllers that tracked their motion and rendered the environment, providing an immersive and interactive experience (Fig. \ref{fig:flowchart}, left). The simulation contained a realistic Sun-Earth-Moon system in which the participant had control over time and their viewing location (Fig. \ref{pic:earthmoonpic}). Students could control the Moon phases by moving forwards and backwards in time with simple button presses or grabbing the Moon to move it in its orbit. Upon entering the simulation, participants were initially placed on top of the Earth’s north pole, but they could change their position to be far above the Earth to provide a more diagrammatic view of the system or to be near the surface of the Earth to view a realistic horizon as the Sun or Moon rises and sets. The participants were given the guiding questions throughout the experience via a virtual clipboard attached to their hand to help students interact with and learn from the environment. The names of the Moon phases were displayed in the environment as the participant moved the Moon. The simulation was created by our team using the Unity game engine for use with the Oculus Rift VR headset.

        \subsubsection*{Desktop Simulation}
The desktop simulation was designed to mimic the VR activity as closely as possible. In the desktop simulation, participants were shown a realistic Sun-Earth-Moon system on a laptop with controls over the camera position and time (Fig. \ref{fig:flowchart}, middle). The environment was created using the same system as the VR simulation (Fig. \ref{pic:earthmoonpic}). Participants could control the Moon phases by going forwards and backwards in time through arrow key presses. They could also navigate around the environment using the mouse and scroll to zoom. The participants were given the guiding questions using a tablet device outside of the simulated environment. The names of the Moon phases were displayed in the environment as the participant moved the Moon. This simulation was created by our team using the Unity game engine. 

        \subsubsection*{Hands-on activity}
The hands-on activity was based on traditional classroom activities about Moon phases~\cite{newbury}. In this activity, the participant's head represented the Earth; the Moon was represented by a small ball held at arm's length; and the Sun was represented by a stationary light (Fig. \ref{fig:flowchart}, right). Participants were asked to rotate counter-clockwise to observe changes in the Moon's phase, as observed in the shadowed portion of the ball's surface. Through this action, real-world complexities arise where the phenomenon may be inaccurately presented. For example, due to the relative distances and sizes between the ball and the participant's head, participants may see eclipses every month. Subtleties such as inclination and procession of the orbits are uncontrollable and missed in the hands-on activity. The participants were given the guiding questions using a tablet as well as the names of the Moon phases.

    \subsection*{Measures}
To evaluate the effects of the different conditions, we assessed student understanding of Moon phases using existing pre-post assessment instruments, their attitudes towards the activities and the environment, and tracked their movements in the simulations.

        \subsubsection*{Moon phases assessment}
The pre- and post-tests each consisted of 14 multiple choice questions about Moon phases and the Moon's motion relative to the Earth, sourced from existing research-validated assessments ~\cite{2002AEdRv...1a..47H,lindell2002,lindell2005,slater2014}. The selected items included ones where the activities were both likely (orbit direction and period) and unlikely (rise and set times) to impact learning. Because the time between the pre- and post-tests were so short, the questions on each test were isomorphic and matched on content, but not identical. The delayed post-test questions were similar to the ones found on the post-test. On all tests, each question had only one correct answer, and the participant's score was the sum of the number of correct answers, with all questions weighted equally. The item-test correlation and item difficulty turned out to be 0.37 and 0.33 for the pre-test and 0.43 and 0.59 for the post-test respectively based on all participant scores.

We also examined learning across conditions on each sub-topic on the assessments by comparing performance on the isomorphic questions. One of the deciding factors in choosing a certain technology as an educational tool is the affordances it offers to teach different aspects of the lesson. For learning about Moon phases, for example, the orbit and rotation periods are not controlled in the hands-on condition, but are constrained to be realistic in the desktop and VR simulations. Keeping the Moon's orbit fixed is a task the participants must preform in the hands-on condition that does not require conscious effort in the other conditions. Differences between the technologies to control or not control certain aspects of the activities, therefore, may lead to differences in participant performance on knowledge questions in different topic areas despite showing the same performance between conditions on the whole exam. Our test contained a question pair that, upon investigation, was not truly the same from pre-test to post-test. Though they referred to the same general topic of why phases occur, they were not isomorphic and the question pair was removed from our analysis. The removal of this question did not significantly alter the results of our analysis. 

        \subsubsection*{Demographic survey}

The demographic survey was provided at the end of the post-test and asked about a variety of participants' characteristics.

            \textbf{Gender:}
Participants were asked their gender with the choices of Male, Female, or Other. For the analysis involving gender as a variable we removed the three participants who chose Other, due to low sample size.

            \textbf{Video game Experience:}
Participants were asked ``On average, how frequently have you played video games over the past three years?" and the choices were daily, weekly, 1-2 times a month, 1-2 times a year, or never. We grouped the choices of daily, weekly, and 1-2 times a month into the category of `having significant video game experience' and the participants who chose 1-2 times a year or never as `not having significant video game experience.'

            \textbf{VR experience:}
Participants were asked ``How much virtual reality experience did you have before you participated in the experiment today?" and the choices were very minimum, moderate, a lot, or none. We grouped the participants who specified moderate or a lot as `having significant VR experience' and the other participants as `not having significant VR experience'.
Only one participant indicated having a lot of VR experience. We maintained three categories of VR experience: none, minimal VR experience, and moderate to high VR experience. 

            \textbf{Academic Major:}
Participants were asked to pick their academic major from a list or write their own. For our analysis we grouped participants according to whether their major was science-focused or non-science-focused. Non-science majors included arts, humanities, economics, social science, communication, and business. Science majors included physics, astronomy, engineering, biology, chemistry, computer science, information science, and math.

        \subsubsection*{Environmental attitudes survey}
We also added an exploratory set of questions to probe participants' socio-scientific attitudes through measurements of their individual differences in environmental attitudes using 15-items selected from the Environmental Attitude Inventory (EAI)~\cite{milfont2010environmental,Dunlap2000}. We suspected that participants in our VR activity might come out with different socio-scientific attitudes, based on similar outcomes found in previous work~\cite{Chang2016, ahn2016experiencing}. In our simulation, we wanted to test the possibility that viewing the planet from the unique vantage point of space (an astronaut-like perspective of the Earth) might have an impact on environmental consciousness~\cite{Riley2008,Poole2010,stepanova2019space}. This effect is attributed to a recognition of the the planet's limited resources~\cite{Dunlap1978}. Specifically, we selected five questions corresponding to each of three of the EAI’s subscales of particular interest to test whether the intervention influenced participants’ environmental movement activism, environmental threat, and human utilization of nature. Sample items were ``I would not want to donate money to support an environmentalist cause,” ``When humans interfere with nature it often produces disastrous consequences (R)” and ``In order to protect the environment, we need economic growth” (1 = Strongly agree to 7 = Strongly disagree), and the modified scale showed sufficiently reliability (Cronbach’s $\alpha=.83$). 

        \subsubsection*{Activity preference}
After completing the post-test, demographic questions, and EAI, participants were invited to try the other two conditions. They then completed a short survey that asked, ``Today, you have experienced three different ways of learning moon phases. Which of the three ways to simulate the moon phases is your favorite method?" Participants could select either ``Demonstration in virtual reality,'' ``Hands-on demonstration,'' or ``On-screen demonstration." They were then asked, ``Please briefly explain, why did you prefer this learning method?" with an open text box answer.

\subsubsection*{Virtual Reality Only Analyses}
Analyses on presence and movement measures will be included in a subsequent publication, as only those in the virtual reality condition answered these questions or had head and hand movements, specifically, tracked.

            \textbf{Presence:}
Fourteen questions from two presence questionnaires \cite{aymerich2012effects,witmer1998measuring} measured participants' sense of spatial presence in the virtual reality activity. Questions drawn from these surveys involved asking participants how much they agreed with the following statements: ``I was really in outer space,” ``I felt surrounded by outer space,” ``I really visited outer space,” and ``The outer space seemed real.”  
The participants assigned to the VR condition were also asked if they had experienced any simulator sickness during the activity and if it was distracting. 

    \textbf{Movement tracking and controller use:}
The X, Y, and Z position and the pitch, yaw, and roll rotation of participants’ head and hands were recorded for the entire session in the VR condition. Movement from timepoint to timepoint may be calculated as the Euclidean distance (mm) between the positions of a tracker at time one and time two. Because the current paper focuses on the comparisons between the three conditions, analysis of the VR specific data, which includes movement, button presses, and presence measures, will be presented in a subsequent publication focused only on participants' experiences in the VR simulation.

    \subsection*{Data analysis}
After all of the data had been collected we cleaned the data set by merging the survey results and verifying participant responses. The full data set was checked for inaccuracies, duplicates, and was translated into the proper form for analysis. Mathematica and R~\cite{R} packages were used for the analyses. All data and code are available through the CISER data archive (\url{ciser.cornell.edu/data/data-archive/})

First, performance on the pre-test, post-test, and delayed post-test were compared between conditions using Analysis of Variance. Effects of condition on environmental attitudes were compared between conditions using ANOVA. Linear regression analyses were used to evaluate effects of other variables on student performance on the post-test, controlling for pre-test score. Variables were selected based on prior literature and included condition, gender, video game experience, and VR experience. Variables were checked for correlation using the ggpairs function from the GGally package in R, recording each variable level as numeric. The only significant correlation was between gender and video game experience, with a correlation coefficient of 0.47. Gender and video game experience were, therefore, analyzed separately in all regression analyses. Variance Inflation Factors (VIF) were also calculated in all regression analyses to measure possible collinearity. Main effects were tested first without interactions, and then regressions with interactions between condition and video game experience, gender, and VR experience were tested next. For the regression analyses, the base variables were a female, non-science major in the hands-on condition, with no VR experience, and not having significant video game experience.

\section*{Results}
We describe our results grouped by overall differences in learning and attitudes based on condition and then by interactions between other variables. 

    \subsection*{Overall learning between conditions}
\begin{figure}[!ht]
\includegraphics[width=0.8\columnwidth]{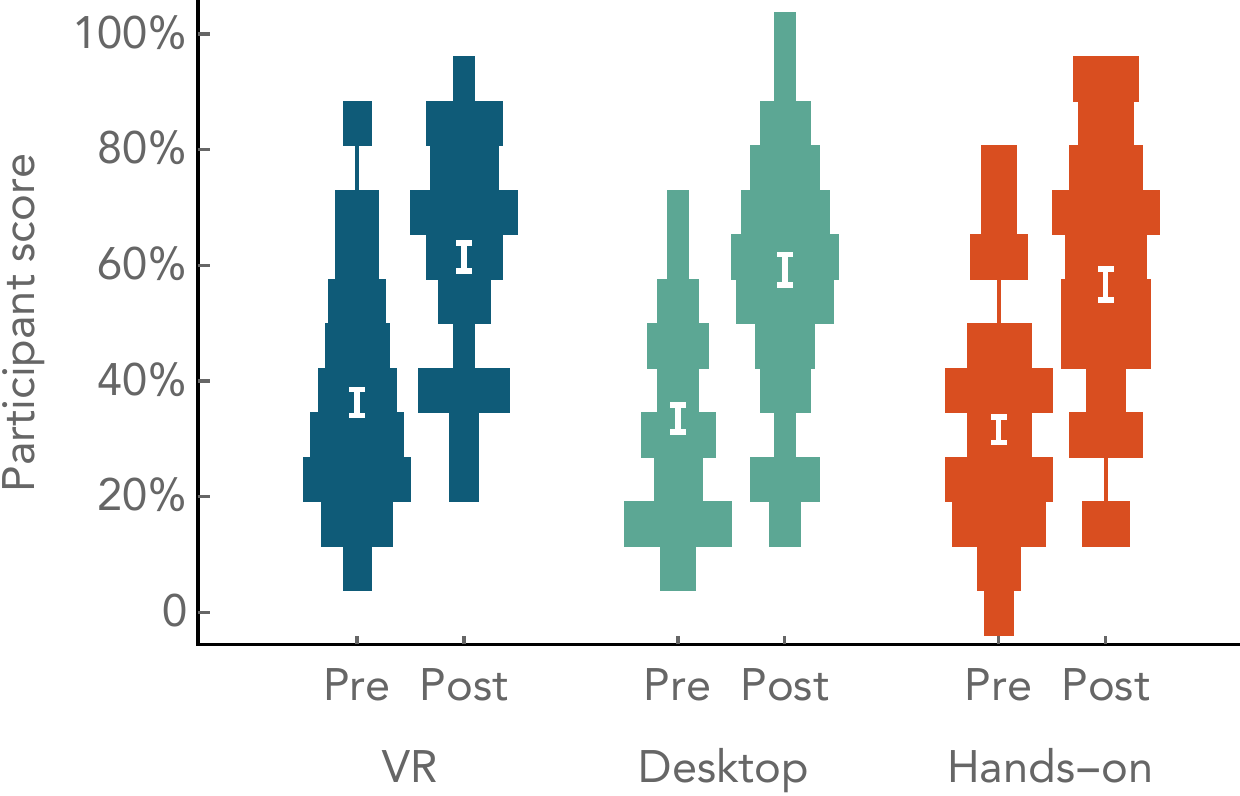}
\caption{{\bf Pre and Post scores by condition}
Figure updated and adapted from~\cite{Madden2018}. An overall view of pre- to post-test performance. Violin plots show how scores were distributed across conditions and between pre- to post-test. Bins are 1 point wide. Average scores and standard error are indicated in white.}
\label{fig:histogram}
\end{figure}

The average score on the pre-test was 33.7\% with no significant differences between the three conditions: \textit{F}(2,169)=1.04, $p=0.356$. Student performance significantly increased from pre- to post-test by 25.3\% on average (average score of 59.1\%), with no significant differences between the post-test scores across the three conditions: \textit{F}(2,169)=0.815, $p=0.444$ (Fig. \ref{fig:histogram}). An ANCOVA controlling for pre-score also showed no differences between the post-score between conditions: \textit{F}(2,168)=1.07, $p=0.344$. The average score on the delayed post-test(completed approximately four months after participation in the activity) was 39.0\%, again with no differences between conditions: \textit{F}(2,51)=0.571, $p=.568$. Our overall normalized gain of 0.38 is similar to that reported in an experiment involving pre-post moon phase assessments for a 20 minute inquiry-based tutorial of 0.54 \cite{lindell2005}.  

Our pre and post tests were not designed to be unidimensional and therefore did not reach a high Cronbach's alpha ($\alpha=0.52$ for pre test, $\alpha=0.66$ for post test). This is expected for multidimensional multiple choice tests \cite{Wilcox2014,Cortina1993}. Both of our tests gave scores across a sufficient range according to Ferguson's delta ($\delta=0.93$ for pre test, $\delta=0.96$ for post test).

\subsection*{Learning by question}
\begin{figure}[!ht]
\includegraphics[width=0.9\columnwidth]{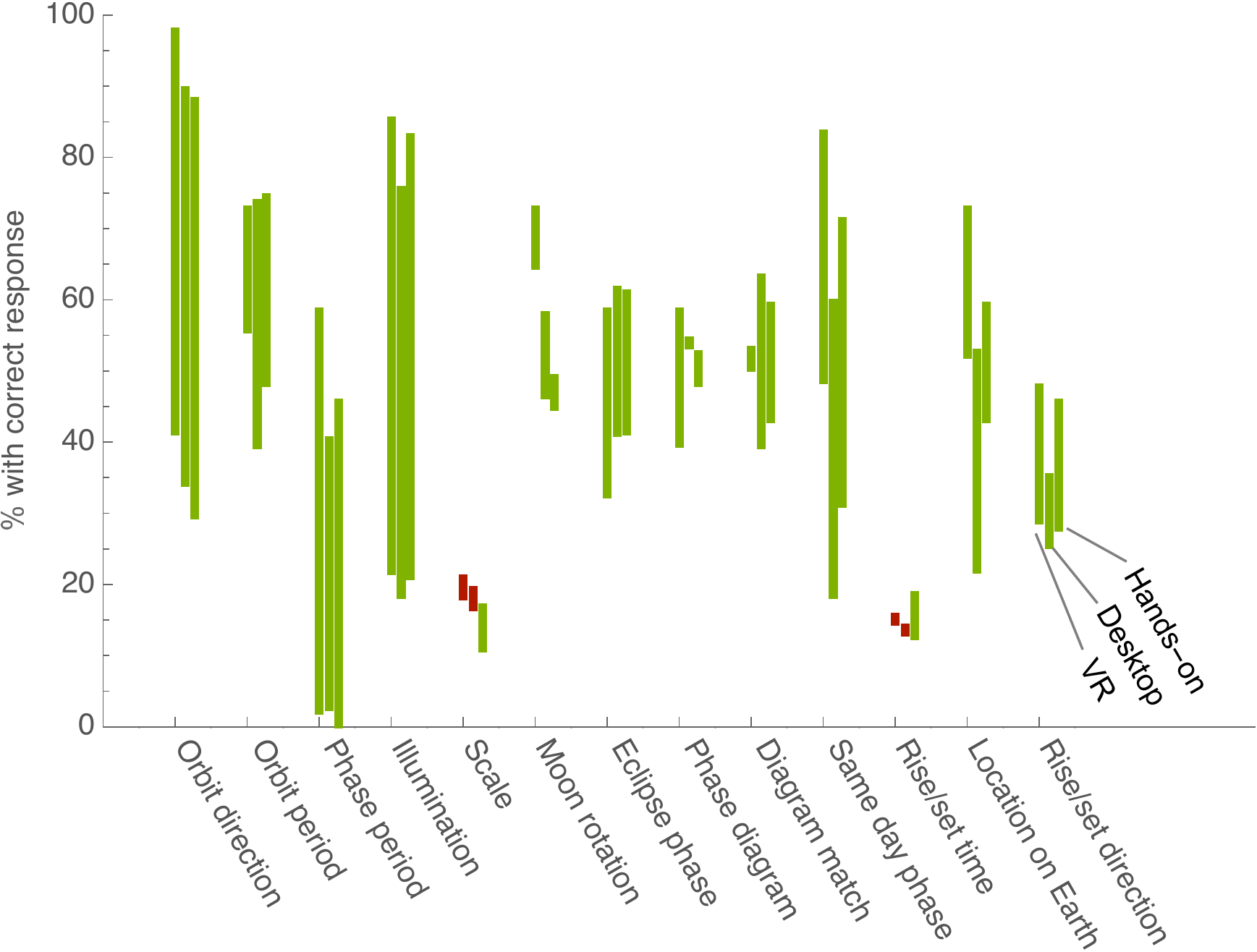}
\caption{{\bf Learning by condition and topic}
Figure updated and adapted from~\cite{Madden2018}. The difference in participant responses from the pre-test to the post-test broken down by question topic. The percent of correct responses on the pre-test and post test for that topic are connected by a colored bar. A green bar signifies improvement, with the higher number representing the post-test score. A red bar means there were fewer correct responses on the post-test, with the higher number representing pre-test score.}
\label{fig:prePostbyQ}
\end{figure}

Breaking down the pre- and post-test scores by question show consistent gains between conditions, with no statistically significant differences (Fig. \ref{fig:prePostbyQ}). Across the conditions, some topic areas showed large gains over 40\% (orbit period, phase period, and illumination), while others showed small gains less than 10\%  (scale, Moon rotation, phase diagram, rise/set time), consistent with previous research on learning about Moon phases ~\cite{lindell2005,wilhelm2018}. This demonstrates that learning did not differ between conditions on sub-topics that may have held learning benefits within the different conditions.

    \subsection*{Attitudes towards the conditions}
\begin{figure}[!ht]
\includegraphics[width=0.4\columnwidth]{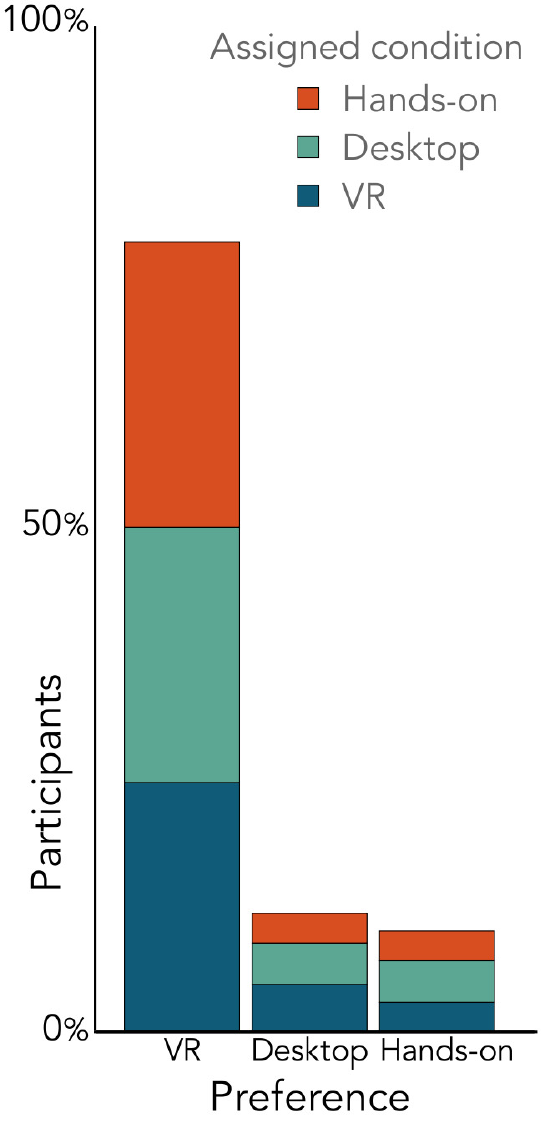}
\caption{{\bf Participant preference}
Preferred choice after viewing each activity. }
\label{fig:preference}
\end{figure}

Consistent with previous work~\cite{dede2009immersive} participants overwhelmingly ($Chi^{2}=152$, $p<0.00001$) preferred the VR condition as their favorite learning method independent of their study condition (Fig. \ref{fig:preference}). 

78\% preferred the VR condition and when describing their reason used phrases such as: the VR condition was ``easier to visualize", ``more realistic", ``more immersive", ``more fun", ``more interesting", and ``the most accurate". The participants who preferred VR generally said that seeing the full picture in a realistic way helped with their learning. The main contributors to this feeling were the ease of viewing different perspectives and having easier control over the system. For example, one participant wrote: 
\begin{quote}
``Having a overall space to see where everything is helps a lot. Even in class I still had a hard time understanding what they are talking about in concept. But I think I learned a lot in VR and being able to manipulate the environment on my own accord. It seems more engaging than the 2 other methods."
\end{quote}

For the few students who did not prefer VR, participants said that they  ``did not notice everything that was going on", found it ``a little too complex", it ``made me dizzy and confused," or generally indicated feeling uncomfortable or overwhelmed. They preferred either the hands-on or desktop conditions because they were more familiar. For example, a participant who preferred the desktop condition wrote:
\begin{quote}
``The VR was cool but since I'm very new to it I spent most of my time just trying to figure out how it worked--it was also tough to find where the Sun and Moon were at times because of how `large' the environment was. The desktop game was more familiar and easy-to-use for me. Personally."
\end{quote}

The 12\% of participants who preferred the desktop condition also used phrases such as: the desktop condition was ``less overwhelming", ``easier to control", and ``very easy to follow". When participants did not like the desktop condition they said: the desktop condition ``gave a limited field of view", and ``[was] a lot harder for me to navigate".

The 10\% of participants who preferred the hands-on condition used phrases such as ``easiest and fastest", ``I was able to more clearly focus", and ``I got distracted by the other methods". A participant who preferred the hands-on condition wrote: 
\begin{quote}
``I really liked the virtual reality method. And it gave me more information than the other two methods, for instance, what time of day certain Moon phases would rise and set. Nevertheless, it was almost too overwhelming and it was as if I was too excited to be in space to actually commit to learning the Moon phases. With the hands-on demonstration. there was nothing to distract me. And, obviously, controlling the demonstration felt about as natural as possible." 
\end{quote}

    \subsection*{Effects on Environmental Attitudes}
Our second hypothesis, that there would be a main effect of condition on environmental attitudes, such that participants in the VR conditions would evince greater environmental support, was also not supported. A one-way ANOVA found no significant differences in environmental attitudes among treatment groups (means $\pm$ standard error were VR = 5.24 $\pm$ 0.09,  Desktop = 5.25 $\pm$ 0.1, Hands-on = 5.35 $\pm$ 0.08; \textit{F}(2, 169) 0.41, p = .66), suggesting that the methods of learning Moon phases did not affect participants’ environmental attitudes. Although we expected that brief exposure in the VR condition to the unique vantage point of the Earth could possibly increase participants’ concerns for environmental issues (by priming the concept of limited resources), the instrument's typical use as a trait-level instrument may make it unsurprising that we did not observe such an effect.

    \subsection*{Effects due to participant demographics}
We conducted exploratory analyses to identify whether other variables were interacting with participants' experiences in the three conditions. Based on prior literature, we focused on gender, major, video game experience, and quantity of VR experience. We first explored main effects alone and then explored interactions between the variables and condition. When examining the role of the demographic variables on performance, we found a correlation between video game experience and gender (column 1 of Fig. \ref{fig:VGGenderCorr}), but no correlation between any of the other variables (Fig. \ref{fig:scitypeexp}). Therefore, video game experience and gender were analyzed in separate regression models.

    \subsubsection*{Main effects between condition and demographic variables}
As shown in Table \ref{table:model12}, students' pre-score was a significant predictor of their post-scores across conditions and, again, there was no significant differences in post-score between conditions. We interpret the regression coefficients as the simple effect size, $\beta$ \cite{baguley2009standardized}. The main effects model also found that there is a significant effect for major, with science majors slightly outperforming non-science majors, even when controlling for pre-score. Neither participants' gender, video game experience, or VR experience had significant main effects.

\begin{table}[!ht]
\begin{adjustwidth}{-2.25in}{0in}
\centering
\caption{
{\bf Linear regression models for gender and video game experience including main effects only.} Regression analysis models for students’ post-score with main effects for pre-score, condition, gender, video game experience, VR experience, and students’ major. Gender and video game experience are analyzed in separate models as the variables were highly correlated.}
\begin{tabular}{|l|c|c|c|c|c|c|c|c|c|c|}
\hline
                     & \multicolumn{5}{c|}{Model 1}                           & \multicolumn{5}{c|}{Model 2}                          \\ \hline
Term                 & B     & SE   & $t$   & $p$          & VIF              & B     & SE   & $t$   & $p$          & VIF             \\ \thickhline
$Intercept$          & 4.52  & 0.53 & 8.45  & \textless{}.001$^{***}$ & 0.0  & 4.42  & 0.54 & 8.26  & \textless{}.001$^{***}$ & 0.0 \\ \hline
$Pre\mhyphen Score$  & 0.50  & 0.08 & 6.65  & \textless{}.001$^{***}$ & 1.1  & 0.50  & 0.08 & 6.62  & \textless{}.001$^{***}$ & 1.1 \\ \hline
$Condition:Desktop$  & -0.05 & 0.41 & -0.13 & 0.898        & 1.4              & 0.00  & 0.40 & 0.00  & 0.998        & 1.4             \\ \hline
$Condition:VR$       & 0.23  & 0.41 & 0.56  & 0.578        & 1.4              & 0.32  & 0.41 & 0.78  & 0.44         & 1.4             \\ \hline
$Gender:Male$        & 0.52  & 0.43 & 1.20  & 0.233        & 1.0              & \multicolumn{5}{c|}{}                                 \\ \hline
$VGexp:Significant$  & \multicolumn{5}{c|}{}                                  & 0.63  & 0.35 & 1.79  & 0.075        & 1.1             \\ \hline
$VRexp:Minimal$      & 0.70  & 0.37 & 1.89  & 0.060        & 1.3              & 0.70  & 0.37 & 1.91  & 0.058        & 1.3             \\ \hline
$VRexp:Moderate$     & -0.17 & 0.51 & -0.33 & 0.741        & 1.4              & -0.19 & 0.50 & -0.39 & 0.699        & 1.4             \\ \hline
$Major:Science$      & 1.14  & 0.36 & 3.21  & 0.002$^{**}$ & 1.1              & 1.00  & 0.35 & 2.83  & 0.005$^{**}$ & 1.2             \\ \thickhline
$R^{2}$              & \multicolumn{5}{c|}{0.325}                      & \multicolumn{5}{c|}{0.330}                                   \\ \hline
$R^{2}_{Adj}$        & \multicolumn{5}{c|}{0.296}                      & \multicolumn{5}{c|}{0.302}                                   \\ \hline
$AIC$                & \multicolumn{5}{c|}{748.41}                     & \multicolumn{5}{c|}{758.67}                                  \\ \hline
\end{tabular}
\begin{flushleft} 
$^* p<.05$, $^{**} p<.01$, $^{***} p<.001$ %Table notes
\end{flushleft}
\label{table:model12}
\end{adjustwidth}
\end{table}

    \subsubsection*{Interactions between condition and demographic variables}
Experience in VR did not have a significant interaction with condition (Table \ref{table:model34}). Notably, this indicates that having experience with VR did not improve students' learning in the VR condition, even though some students claimed that not having experience in VR impeded their learning.

\begin{figure}[!ht]
\includegraphics[width=0.8\columnwidth]{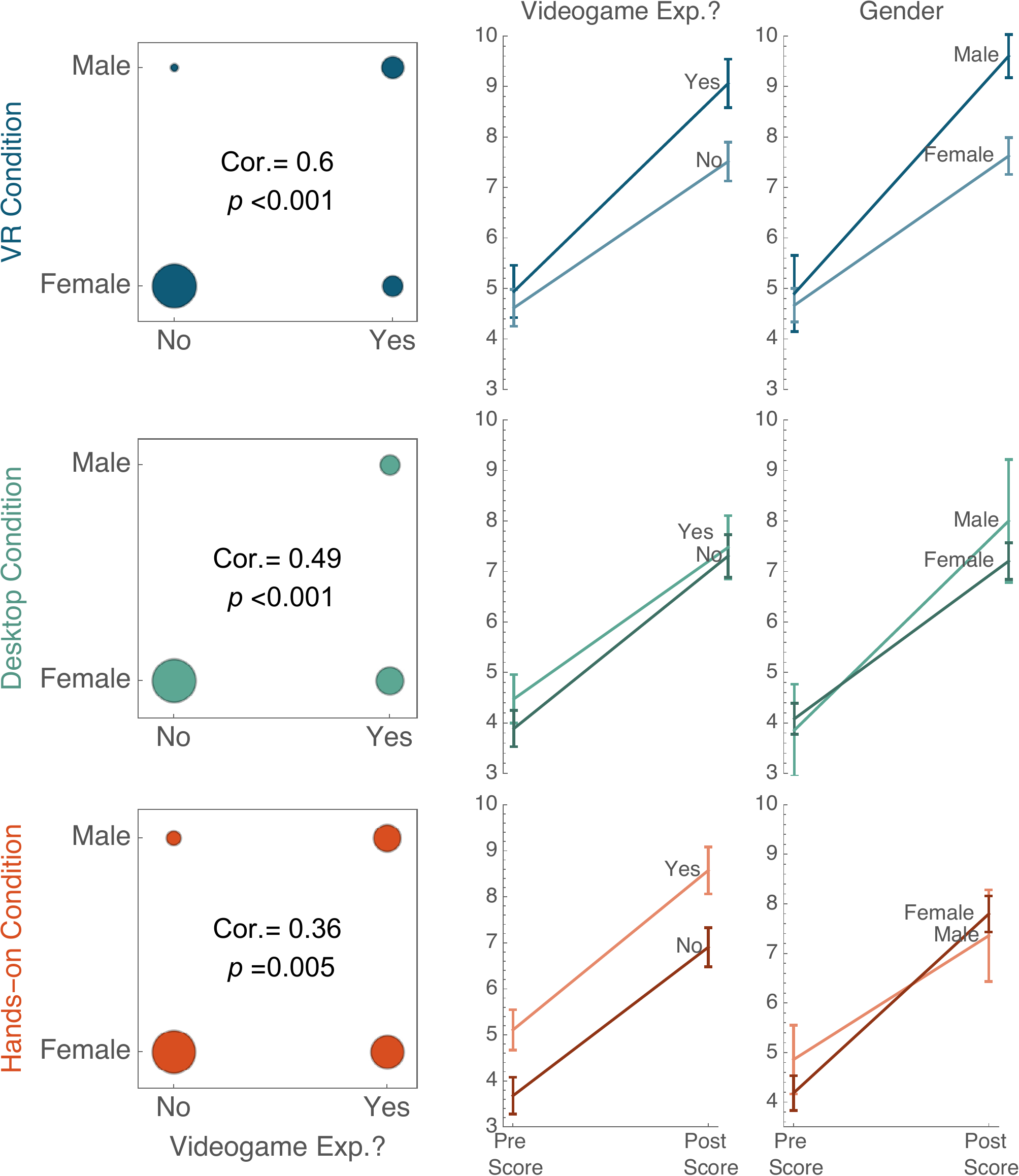}
\caption{{\bf Relationship between gender, video game experience, and scores across conditions}
The first column shows the correlations between gender and video game experience. The second column shows the average scores and standard error at pre- and post-test for students with and without video game experience. The third column shows the average scores and standard error at pre- and post-test for students identifying as male and female. See S1 Table for line details.}
\label{fig:VGGenderCorr}
\end{figure}

The interactions with gender and video game experience are more complicated. As seen in Table \ref{table:model34}, there is an interaction between gender and condition in Model 3, with men outperforming in the VR condition. However, Model 4 indicates that there is a main effect with video game experience (seen as marginally significant in Model 2), and a potential interaction between condition and video game experience. From Fig. \ref{fig:VGGenderCorr}, this interaction is likely understood through the pre-test scores. Only in the hands-on condition do we see differences in pre-test scores between participants with and without video game experience. We believe these results can be understood such that men and participants with video game experience learn more from the VR condition, followed by the desktop and then hands-on conditions. Given the correlation between video game experience and gender, it is unclear which variable is responsible for the differences. This correlation between gender and video game experience may be more clearly understood by elaborating on the type and detailed quantity of video game experience in future studies. 

\begin{table}[!ht]
\begin{adjustwidth}{-2.25in}{0in}
\centering
\caption{
{\bf Linear regression models including main effects with gender and video game interaction.} Regression analysis models for students’ post-score with main effects for pre-score, condition, gender, video game experience, VR experience, and students’ major and interactions between condition and gender, VR experience, and video game experience. Gender and video game experience are analyzed in separate models as the variables were highly correlated.}
\begin{tabular}{|l|c|c|c|c|c|c|c|c|c|c|}
\hline
                                         & \multicolumn{5}{c|}{Model 3}                       & \multicolumn{5}{c|}{Model 4}                        \\ \hline
Term                                     & B    & SE   & $t$  & $p$ & VIF                     & B    & SE   & $t$  & $p$ & VIF                      \\ \thickhline
$Intercept$                              & 4.76 & 0.61 & 7.74 & \textless{}.001$^{***}$& 0.0 & 4.15 & 0.65 & 6.38 & \textless{}.001$^{***}$ & 0.0 \\ \hline
$Pre\mhyphen Score$                      & 0.50 & 0.08 & 6.71 & \textless{}.001$^{***}$& 1.1 & 0.48 & 0.08 & 6.28 & \textless{}.001$^{***}$ & 1.1 \\ \hline
$Condition:Desktop$                      & -0.55& 0.72 & -0.77& 0.443 & 4.4                   & 0.83 & 0.79 & 1.05 & 0.296 & 5.7                    \\ \hline
$Condition:VR$                           & -0.25& 0.68 & -0.36& 0.719 & 4.0                   & 0.55 & 0.75 & 0.73 & 0.465 & 5.1                    \\ \hline
$Gender:Male$                            & -0.76& 0.67 & -1.14& 0.258 & 2.6                   & \multicolumn{5}{c|}{}                               \\ \hline
$VGexp:Significant$                      & \multicolumn{5}{c|}{}                              & 1.42 & 0.58 & 2.43 & 0.016$^{*}$ & 3.3              \\ \hline
$VRexp:Minimal$                          & 0.88 & 0.62 & 1.41 & 0.160 & 3.7                   & 0.92 & 0.61 & 1.50 & 0.136 & 3.8                    \\ \hline
$VRexp:Moderate$                         & -0.42& 0.88 & -0.48& 0.634 & 4.3                   & -1.11& 0.87 & -1.27& 0.205 & 4.2                    \\ \hline
$Major:Science$                          & 1.18 & 0.35 & 3.37 & 0.001$^{**}$ & 1.1            & 1.06 & 0.36 & 2.99 & 0.003$^{**}$ & 1.2             \\ \hline
$Gender:Male \times Condition:Desktop$   & 1.77 & 1.12 & 1.59 & 0.115 & 1.8                   & \multicolumn{5}{c|}{}                               \\ \hline
$Gender:Male \times Condition:VR$        & 2.56 & 1.00 & 2.57 & 0.011$^{*}$ & 2.2             & \multicolumn{5}{c|}{}                               \\ \hline
$VRexp:Minimal \times Condition:Desktop$ & 0.28 & 0.89 & 0.32 & 0.752 & 4.3                   & -0.17& 0.87 & -0.20& 0.841 & 4.4                    \\ \hline
$VRexp:Minimal \times Condition:VR$      &-0.66 & 0.90 & -0.73& 0.467 & 3.4                   & -0.74& 0.90 & -0.83& 0.407 & 3.5                    \\ \hline
$VRexp:Moderate \times Condition:Desktop$&-0.18 & 1.29 & 0.14 & 0.888 & 2.5                   & -0.09& 1.28 & -0.07& 0.943 & 2.5                    \\ \hline
$VRexp:Moderate \times Condition:VR$     & 0.95 & 1.15 & 0.83 & 0.409 & 3.6                   & 1.78 & 1.14 & 1.56 & 0.120 &  3.5                   \\ \hline
$VG exp:Has \times Condition:Desktop$    & \multicolumn{5}{c|}{}                              & -1.82& 0.84 & -2.17& 0.03$^{*}$ & 3.0               \\ \hline
$VG exp:Has \times Condition:VR$         & \multicolumn{5}{c|}{}                              & -0.40& 0.86 & -0.47& 0.642 & 2.7                    \\ \thickhline
$R^{2}$                                  & \multicolumn{5}{c|}{0.373}                         & \multicolumn{5}{c|}{0.372}                          \\ \hline
$R^{2}_{Adj}$                            & \multicolumn{5}{c|}{0.321}                         & \multicolumn{5}{c|}{0.321}                          \\ \hline
$AIC$                                    & \multicolumn{5}{c|}{747.89}                        & \multicolumn{5}{c|}{759.59}                         \\ \hline
\end{tabular}
\begin{flushleft} 
$^* p<.05$, $^{**} p<.01$, $^{***} p<.001$ %Table notes
\end{flushleft}
\label{table:model34}
\end{adjustwidth}
\end{table}

\begin{figure}[!ht]
\includegraphics[width=0.5\columnwidth]{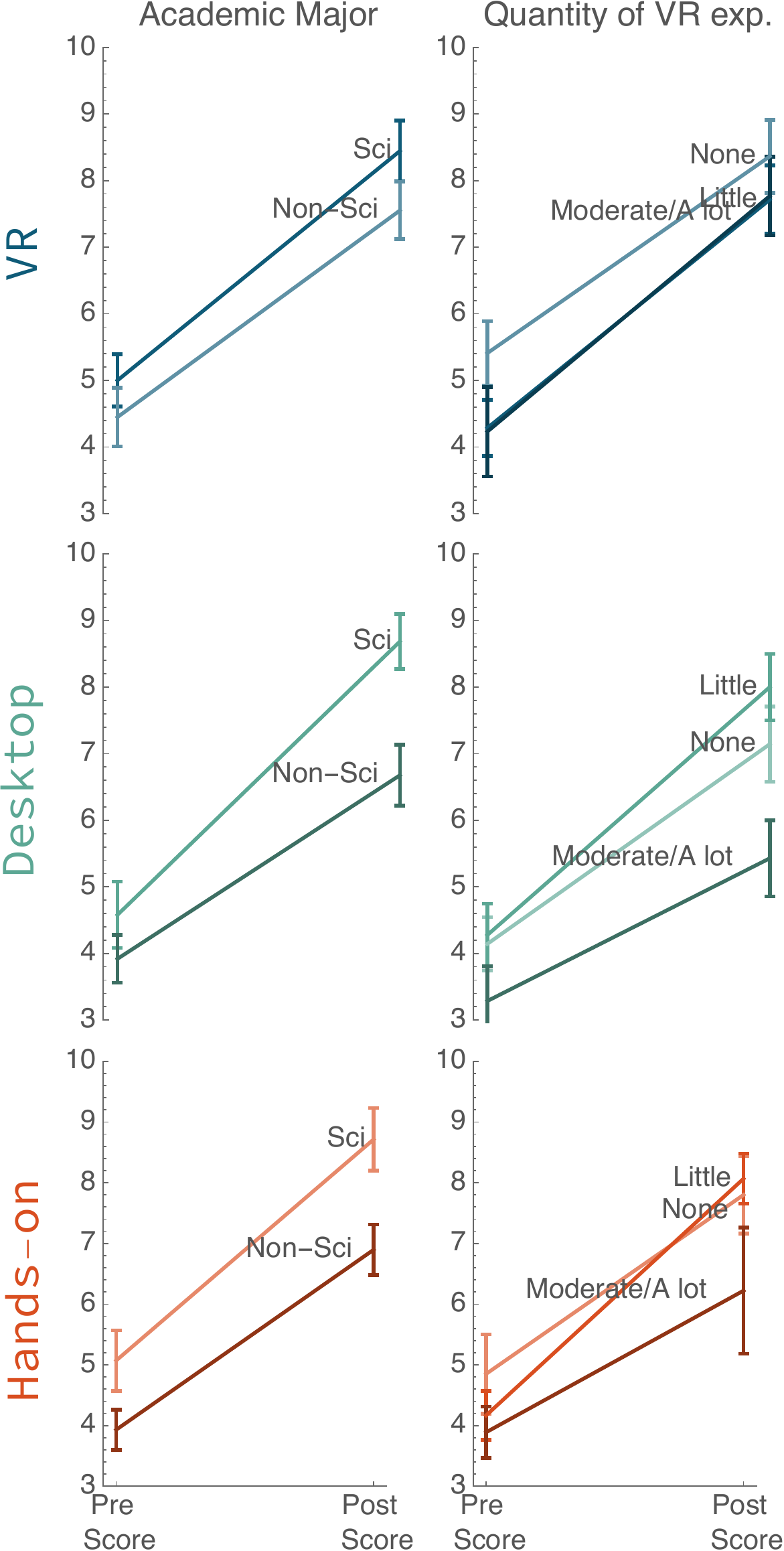}
\caption{{\bf Interactions plots for academic major, and quantity of VR experience}
Each column shows the relationship present between one of these measures and pre to post score by condition. Mean score and standard error are shown, see S2 Table for line details.}
\label{fig:scitypeexp}
\end{figure}

\section*{Discussion}
In this study, we performed a controlled experiment of student learning about Moon phases through three different modalities: a hands-on activity, a desktop simulation, and a VR simulation. We found no overall effect for condition on students' learning on an immediate post-test, nor on a delayed post-test four months after the intervention. We did find that students with declared science majors outperformed non-science majors in all conditions. Student learning in the VR condition was not improved with VR experience. Despite these results, students overwhelmingly preferred learning in the VR condition. 

There are several possible interpretations of these results. First, our hypothesis had been that the VR simulation would improve performance by reducing the real-world complexities of a hands-on activity and providing a realistic, embodied learning experience. The results suggest that these affordances did not impact learning about Moon phases. On the other hand, one may interpret that the VR condition provided equivalent learning to the other two modalities, while dramatically improving students' attitudes towards the learning experience. 

The study also explored interactions between conditions and students' video game experience and gender. Gender and video game experience were significantly correlated in our study, and men performed better in the VR condition. This means that either video game experience, being male, or a combination of the two provides an advantage. 

There are several studies that suggest video game experience would provide this advantage. One study found that video games can have a beneficial effect on completing complex spatial tasks, visuomotor coordination, and multiple object tracking\cite{Spence2010}. Other studies suggest that the benefits of video games on such abilities can be gained after a short time playing an action game and is not dependent on gender~\cite{Feng2007}.  Furthermore, research has shown that men are more drawn to the type of video games suggested to provide these benefits~\cite{Phan2012}. Combining these studies paints a picture that these advantages are not inherently male but may be caused by a particular type of video game experience that so happens to be common for men. 

Alternatively, since the 1970's, several studies have suggested that men have better spatial reasoning than women. However, comprehensive studies have shown recently that, while differences in spatial reasoning between men and women may be present in certain cases, we may not fully understand their causes. Newcombe and Stieff claimed, ``from a practical educational standpoint, the most relevant fact is that the relevant skills can be improved in both men and women"~\cite[p. 962]{Newcombe2012}. Research has also previously found that men outperformed women on post-test conceptual assessments of Moon phase understanding, but only on items involving spatial reasoning~\cite{Wilhelm2009}. However, it has also been found that men's and women's scores improve similarly from pre- to post-test with appropriate instruction~\cite{Wilhelm2009, Jackson2015}, again suggesting that performance can be improved in both men and women. 

While our study was unable to prove video game experience was the true contributor to performance increases in the VR condition over gender, literature on the subject suggests that video game experience and not gender may be the affecting variable. While further experimental work should examine this proposition, this points the way to improving learning experiences in virtual reality such that they benefit all learners. If all participants did as well in the virtual reality condition as did men/people with video game experience, then virtual reality could be an overall more effective teaching tool than the other two tested modes of teaching moon phases. 

    \subsection*{Limitations}
There are several limitations to this study that should be considered when interpreting our results. This study was focused on learning in an activity that was based on physics and astronomy concepts. Thus, we should be careful when applying our findings to learning in other subjects or even other concepts within physics and astronomy \cite{Winn2002}. Furthermore, the participant pool for this study was not perfectly representative of either a typical college classroom or typical physics/astronomy classroom. For example, our study was comprised of 80\% women while the American Physical Society reports that only around 20\% of undergraduate physics degrees are awarded to women~\cite{APSGender}. In addition, the significant interaction between gender and condition was based on a small sample of male participants and future work should evaluate this result with more equally distributed samples.

The contribution that we believe this paper makes to the discussion on gender/video game effects on learning in virtual reality is based on exploratory analyses. Thus, future work should explicitly test these hypotheses in a pre-registered study and include more detailed questions related to the type and quantity of video game experience.  

While VR technology has advanced rapidly, it is still not ideal. Control responsiveness, motion sickness, limited resolution and field of view are all technological obstacles that can still break immersion and distract from learning using today's equipment. In contrast, the technologies we used for the hand-on condition and the desktop condition have essentially plateaued compared to VR, meaning the results of this study may change as VR technology progresses. 

Compared to VR, the participants were very familiar with the technology used in the hands-on condition and desktop condition. Many participants had never experienced VR before our study or had very limited experience with using a VR headset. This suggests that many of our participants were managing high cognitive load as they attempt to become comfortable with VR, understand the activity, and learn about Moon phases. Participants in the other two conditions did not have such a high barrier to getting comfortable with the technology handed to them. For participants who preferred the non-VR conditions, a common praise was familiarity with the equipment. This suggests that as a college population becomes more familiarly and comfortable using VR, the results of this study may change. Future research should measure students' cognitive load during the activity explicitly, such as through eye-tracking~\cite{Zu2018}, self-report surveys~\cite{Zu2018, Paas1992, Leppink2013, Paas2003, Paas1994}, physiological indicators~\cite{Paas1994, Paas2003}, or electroencephalography~\cite{Antonenko2010}. We note, however, that many of the common methods have several limitations~\cite{DeJong2010}, such as that different methods may or may not be able to distinguish different types of cognitive load~\cite{Leppink2013, DeLeeuw2008, DeJong2010, Paas2003}. It is unclear which may be at play in these activities.

\subsection*{Future Work}

Our study joins others in suggesting that virtual reality is a promising technology as an educational tool but does not, in itself, guarantee a learning advantage over traditional hands-on activities or desktop simulations. There are several areas for future work, based on this analysis.

Future work should focus on further investigating the potential relationship between video game experience and learning gains. This strategy has three components. First, researchers should attempt to sample both male and female participants with equivalent video game experience. Video game experience should be more precisely characterized, focusing on game types that require more visuospatial navigation skills. Video game experiences tend to build procedural or motor skills \cite{rosser2007impact} which do not typically go away after long-term disuse. Measures should thus include lifetime experience, rather than frequency of use alone. If video game experience is confirmed to provide users with skills that allow them to learn better in VR, then researchers can use this knowledge to provide participants with tools to learn to interact with the environment in order to provide a level playing field to all participants. 

The VR educational experience can also be improved through enhancing the user interface to maximize VR’s full potential. This does not require waiting for the technology to improve or to become more widely used; instead, designers can take cues from participant's responses. One strategy would be to enhance users' sense of embodiment. In this study, users were not embodied in avatars, which may have affected their feeling of presence or ‘being there’ within an environment \cite{slater2017implicit}. This in turn could have affected how willing participants were to interact with the environment, thereby reducing their learning gains. Examining existing movement data in the VR condition could provide hints for how to design such a simulation. Similarly, our study used guiding questions that must be answered before proceeding into the environment which may have prevented exploration. Thus, future studies should evaluate how avatar embodiment and interactive text affect movement and learning within an environment as well as how visuospatial ability relates to movement in VR. 

Finally, participants worked alone in our study, but each of our conditions could also be used to collaborate. Two students could help each other learn concepts through answering questions together, or an instructor could emphasize key components that are easily missed~\cite{Chi2009}. Indeed in virtual reality in particular, the anonymity can  make it a safer space for learning \cite{yu2009creating}, thereby encouraging students to make mistakes and learn from them without fear of judgement. If participants experience gains from social learning, such gains may be more noticeable in virtual environments.

However, future work must also remain open to the possibility that the excitement and engagement produced by virtual reality experiences may not translate into learning gains in all domains.

\section*{Conclusion}
This study has several takeaway findings. First, participants' learning gains from pre- to post-test were not significantly different, on average, between the VR, desktop, and hands-on conditions. Participants preformed similarly well on each question topic across the three conditions. We found no strong evidence that participants' retention differed between conditions after four months. Our hypothesis, that VR would improve learning by simplifying real-world complexities and providing an embodied learning experience, was not supported. Nonetheless, participants strongly favored learning in the VR activity. 

Guided by the literature on virtual reality and learning, we also collected data on demographic measures (gender, video game experience, virtual reality experience, and major) to explore predicted interactions, all of which are reported here. We did find a positive effect from gender within the VR condition. However, video game experience and gender were significantly correlated in our study, and the literature suggests that video game experience may be the main reason for the performance increase. 

This study has allowed us to determine new experiment designs that will help explore the reasons we saw similar learning gains across conditions. What remains promising about VR is that, relative to a ball on a stick and 2D computer games, it is a technology that is rapidly advancing. Given that participants' unfamiliarity with VR and the technical roughness of the simulation, the fact that participants were able to learn as much as those in the other conditions may bode well to support VR as a better educational tool when the majority of students are comfortable learning in a virtual environment.

Given that the learning was the same regardless of condition what remains is the fact that participants widely favored the VR experience. As a method of engaging students, using VR was successful in our study and was not achieved at the cost of learning gains. Novelty does diminish however, so an advantage based on novelty alone will cease to be an advantage as exposure rises. 

All together, there are many avenues to explore with educational VR. Future work should encourage a more comprehensive look into VR’s ability as an educational tool, such that a participant’s experience is considered from multiple perspectives. 

\section*{Acknowledgments}
This work was supported by Oculus Education with special thanks to Cindy Ball and Cindi McCurdy. We would like to thank all the graduate and undergraduate students who helped develop the simulations and run participants: Tristan Stone, Yilu Sun, Akhil Gopu, Kristi Lin, Jason Wu, Anirudh Maddula, Frank Rodriguez, Alice Nam, Phil Barrett, Dwyer Tschantz, Connor Lapresi, Giulia Reversi, Keun Youk, Albert Tsao, and Annie Hughey. We would also like to thank Kimberly Williams from the Cornell Center for the Integration of Research, Teaching, and Learning, Stephen Parry of the Cornell Statistical Consulting Unit, Daniel Alexander and Florio Arguillas from the Cornell Institute of Social and Economic Research, and the entire Cornell Physics Education Research Lab for support, comments, and assistance. 

\nolinenumbers

\newpage
% Goes before acknowledgements. For more information, see \nameref{S1_Appendix}.
\section*{Supporting information}
% Include only the SI item label in the paragraph heading. Use the \nameref{label} command to cite SI items in the text.
\paragraph*{S1 Table}
{\bf Modeling interactions between video game experience and gender with condition} Supporting data for Fig. \ref{fig:VGGenderCorr}
\begin{table}[!ht]
\begin{adjustwidth}{-2.25in}{0in} % Comment out/remove adjustwidth environment if table fits in text column.
\centering
\begin{tabular}{|c|c|c|c|c|c|c|c|}
\thickhline
\multicolumn{8}{|c|}{Video game experience}                               \\ \hline
Condition & Exp.   & Intercept & SE   & p       & Slope & SE   & p       \\ \hline
VR        & Yes    & 4.94      & 0.50 & 3.0E-11 & \textbf{4.12}  & 0.71 & 1.8E-06 \\ \hline
VR        & No     & 4.62      & 0.37 & 7.1E-20 & \textbf{2.90}  & 0.53 & 5.1E-07 \\ \hline
Desktop   & Yes    & 4.48      & 0.56 & 7.7E-10 & 3.00  & 0.79 & 4.9E-04 \\ \hline
Desktop   & No     & 3.89      & 0.39 & 5.5E-15 & 3.42  & 0.55 & 4.0E-08 \\ \hline
Hands-on  & Yes    & \textbf{5.11}      & 0.48 & 5.6E-15 & 3.46  & 0.67 & 3.9E-06 \\ \hline
Hands-on  & No     & \textbf{3.68}      & 0.41 & 1.4E-12 & 3.23  & 0.58 & 7.6E-07 \\ \thickhline
\multicolumn{8}{|c|}{Gender}                                             \\ \hline
Condition & Gender & Intercept & SE   & p       & Slope & SE   & p       \\ \hline
VR        & Male   & 4.90      & 0.61 & 2.4E-07 & \textbf{4.70}  & 0.86 & 3.7E-05 \\ \hline
VR        & Female & 4.67      & 0.35 & 5.2E-23 & \textbf{2.96}  & 0.49 & 4.1E-08 \\ \hline
Desktop   & Male   & 3.86      & 1.07 & 3.7E-03 & 4.14  & 1.52 & 1.8E-02 \\ \hline
Desktop   & Female & 4.08      & 0.34 & 4.2E-21 & 3.12  & 0.48 & 2.6E-09 \\ \hline
Hands-on  & Male   & 4.86      & 0.82 & 2.8E-06 & 2.50  & 1.16 & 4.0E-02 \\ \hline
Hands-on  & Female & 4.18      & 0.36 & 1.6E-19 & 3.61  & 0.50 & 2.5E-10 \\ \hline
\end{tabular}
\begin{flushleft} %Table notes
\end{flushleft}
%\label{table:VG%Gender}
\end{adjustwidth}
\end{table}
%\newpage
\paragraph*{S2 Table}
{\bf Modeling interactions between academic major, and VR experience quantity with condition} Supporting data for Fig. \ref{fig:scitypeexp}
\begin{table}[!ht]
\begin{adjustwidth}{-2.25in}{0in} % Comment out/remove adjustwidth environment if table fits in text column.
\centering
\begin{tabular}{|c|c|c|c|c|c|c|c|}
\thickhline
\multicolumn{8}{|c|}{Academic Major}                                             \\ \hline
Condition & Major          & Intercept & SE   & p       & Slope & SE   & p       \\ \hline
VR        & Science        & 5.40      & 0.62 & 1.6E-09 & 3.27  & 0.87 & 8.1E-04 \\ \hline
VR        & Non-Science    & 5.43      & 0.99 & 1.5E-04 & 2.29  & 1.41 & 1.3E-01 \\ \hline
Desktop   & Science        & 4.38      & 0.37 & 1.2E-08 & \textbf{3.88}  & 0.52 & 3.4E-06 \\ \hline
Desktop   & Non-Science    & 4.17      & 0.79 & 2.7E-05 & \textbf{2.17}  & 1.12 & 6.5E-02 \\ \hline
Hands-on  & Science        & \textbf{5.75}      & 0.89 & 1.7E-06 & 2.83  & 1.26 & 3.5E-02 \\ \hline
Hands-on  & Non-Science    & \textbf{4.00}      & 0.72 & 1.3E-04 & 3.00  & 1.02 & 1.3E-02 \\ \thickhline
\multicolumn{8}{|c|}{Quantity of VR experience}                                  \\ \hline
Condition & Quantity       & Intercept & SE   & p       & Slope & SE   & p       \\ \hline
VR        & None           & \textbf{5.41}      & 0.52 & 2.8E-13 & 2.95  & 0.73 & 2.2E-04 \\ \hline
VR        & Little         & 4.29      & 0.47 & 2.4E-11 & 3.43  & 0.66 & 6.8E-06 \\ \hline
VR        & Moderate/A lot & 4.23      & 0.63 & 6.3E-07 & 3.54  & 0.89 & 5.8E-04 \\ \hline
Desktop   & None           & 4.14      & 0.49 & 2.1E-10 & 3.00  & 0.70 & 1.0E-04 \\ \hline
Desktop   & Little         & 4.28      & 0.48 & 3.5E-12 & 3.72  & 0.69 & 1.3E-06 \\ \hline
Desktop   & Moderate/A lot & 3.29      & 0.55 & 6.2E-05 & \textbf{2.14}  & 0.77 & \textbf{1.7E-02} \\ \hline
Hands-on  & None           & 4.85      & 0.65 & 5.2E-09 & 2.95  & 0.91 & 2.6E-03 \\ \hline
Hands-on  & Little         & 4.17      & 0.41 & 1.7E-14 & 3.90  & 0.58 & 8.3E-09 \\ \hline
Hands-on  & Moderate/A lot & 3.89      & 0.79 & 1.6E-04 & 2.33  & 1.12 & 5.4E-02 \\ \hline
\end{tabular}
\begin{flushleft} %Table notes
\end{flushleft}
%\label{table:VG%Gender}
\end{adjustwidth}
\end{table}
\newpage

%\bibliography{MoonPhases}

\providecommand{\noopsort}[1]{}\providecommand{\singleletter}[1]{#1}%

%\section*{Supporting information}
%
%\paragraph*{S1 Table}
%{\bf Modeling interactions between video game experience and gender with condition} Supporting data for Fig. \ref{fig:VGGenderCorr}
%
%
%\paragraph*{S2 Table}
%{\bf Modeling interactions between academic major, and VR experience quantity with condition} Supporting data for Fig. \ref{fig:scitypeexp}
%

\end{document}